\documentclass{aa} 
\usepackage{graphicx}
\usepackage{epsfig}
\usepackage{natbib}
\bibpunct{(}{)}{;}{a}{}{,}

\begin{document}


\title{Statistical properties of exoplanets.}  \subtitle{I. The period
  distribution: constraints for the migration scenario}

\author{ S.~Udry \and M.~Mayor \and N.C. Santos}

\offprints{S. Udry, \email{stephane.udry@obs.unige.ch}}

\institute{Observatoire de Gen\`eve, 51 ch.  des Maillettes, CH--1290
  Sauverny, Switzerland }

\date{Received 4 November 2002 / Accepted 16 May 2003 } 

\abstract{ Interesting emerging observational properties of the
  period-mass distribution of extra-solar planets are discussed.  New
  recent detections confirm the already emphasized lack of massive
  planets ($m_2\sin{i}\ge 2$\,M$_{\rm Jup}$) on short-period orbits
  ($P\le 100$\,days). Furthermore, we point out i) a shortage of
  planets in the 10--100~day period range as well as ii) a lack of
  light planets ($m_2\sin{i}\le 0.75$\,M$_{\rm Jup}$) on orbits with
  periods larger than $\sim$\,100~days. The latter feature is shown
  not to be due to small-number statistics with Monte-Carlo
  simulations. These observational period-related characteristics are
  discussed in the context of the migration process of exoplanets.
  They are found to be in agreement with recent simulations of planet
  interactions with viscous disks. The observed {\sl valley} at a few
  tens of days in the period distribution is interpreted as a
  transition region between two categories of planets that suffered
  different migration scenarios.  The lack of light planets on
  longer-period orbits and the corresponding intriguing sharp limit in
  mass is tentatively explained by the runaway migration process
  recently studied by Masset \& Papaloizou (2003).  The observed
  properties also have implications for the observation strategies of
  the on-going surveys and of future higher-precision searches.
   }
\maketitle

\section{Introduction}

The most remarkable feature of the sample of known extra-solar planets
is undoubtedly the variety of their orbital characteristics, which
challenges the conventional views of planetary formation. Amongst the
most peculiar candidates are the giant planets orbiting very close to
their parent stars, in contrast to the prediction of the {\sl standard
  model} \citep[e.g.][]{Pollack-96} that they formed first from ice
grains\footnote{Such grain growth provides the supposed requisite
  solid core around which gas could rapidly accrete, over the lifetime
  of the protoplanetary disk ($\sim10^7$\,y)} in the outer region of
the system where the temperature of the stellar nebula is not too
high.  The accommodation of this scenario to the present observations
requires that the planets undergo a subsequent migration process
bringing them close to the central star \citep[see
e.g.][]{Lin-96,Ward-97}.  Alternative points of view invoke {\sl
  in-situ} formation \citep{Wuchterl-2000:a,Wuchterl-2000:b},
possibly triggered through disk instabilities
\citep{Boss-2001,Boss-2002}.  Note however that, even in such
cases, subsequent disk-planet interactions leading to migration are
expected to take place as soon as the planet is formed.

The number of known extra-solar planets exceeding 100, a statistically
significant sample is now available from which we can determine
meaningful distributions of planetary characteristics
\citep{Udry-2003:a,Marcy-2003}, and so try to point out useful
constraints for the planet-building models and then possibly
discriminate between the different proposed scenarios.

In a new series of papers, we will try to emphasize the emerging
properties of planet-host stars and characteristics of the different
orbital-element distributions of exoplanetary systems and discuss
their implications for our understanding of planetary formation and
evolution.  This paper (Paper\,I) will be dedicated to the
period/separation distribution of exoplanets.  Sharp features in the
period versus mass diagram start to emerge and provide strong
observational constraints for the migration scenario. More
specifically, we re-discuss the lack of massive planets on
short-period orbits recently pointed out
\citep{Udry-2002,Zucker-2002,Patzold-2002}. We also emphasize the
clear emergence of a planet shortage in the 10--100\,d period range (a
{\sl period valley}) and we point out a sharp lower mass limit for the
detected planets on ``longer''-period orbits ($P\ge 100$\,d). The later
part of the paper will be dedicated to a discussion of the proposed
explanations of these important observational findings and the
possible implications for the future surveys.

\begin{figure*}[t]
\centerline{\psfig{width=0.75\hsize,file=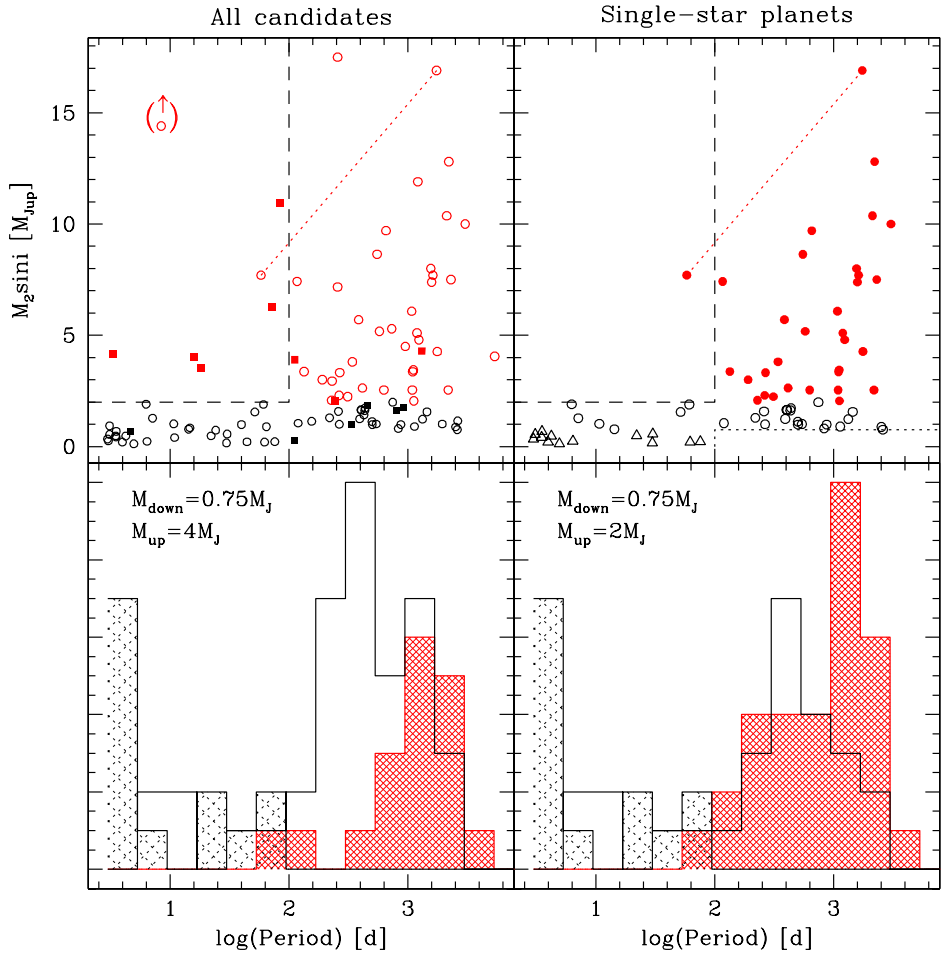}}
\caption{
\label{fig1}
{\sl Upper panels:} Minimum masses versus periods for known exoplanet
candidates. In the left panel, filled squares indicate planets in
binaries \citep[Paper\,III;][]{Udry-2003:b,Udry-2003:c} whereas
circles are used for planets around single stars. The point in "()"
represents {\object HD\,162020} for which synchronisation arguments
indicate a probable mass in the brown-dwarf regime \citep{Udry-2002}.
In the right panel, only planets orbiting single dwarf stars are
represented (i.e.  planets in binaries and planets orbiting evolved
stars have been discarded). A different coding is used for massive
($m_2\sin{i}$\,$\geq$\,2\,M$_{\rm Jup}$; filled symbols),
intermediate-mass ($m_2\sin{i}$ between 0.75 and 2\,M$_{\rm Jup}$;
open circles), and lighter ($m_2\sin{i}$\,$\leq$\,0.75\,M$_{\rm Jup}$;
open triangles) candidates.  The components of the possible multi
brown-dwarf system {\object HD\,168443} are linked by a dotted line.
The dashed and dotted lines in the panels indicate limits at
$P$\,=\,100\,d (vertical), at $m_2\sin{i}$\,=\,2\,M$_{\rm Jup}$
(horizontal left), or at $m_2\sin{i}$\,=\,0.75\,M$_{\rm Jup}$
(horizontal right).  {\sl Lower panels:} Period distributions of the
planets orbiting single dwarf stars (shown in the upper-right panel)
for different mass regimes: light-shaded histograms are for the
lightest planets ($m_2\sin{i}$\,$\leq$\,M$_{\rm
  down}$\,=\,0.75\,M$_{\rm Jup}$), open histograms are for
intermediate masses (between M$_{\rm down}$ and M$_{\rm up}$ -- right:
M$_{\rm up}$\,=\,2\,M$_{\rm Jup}$; left: M$_{\rm up}$\,=\,4\,M$_{\rm
  Jup}$), whereas the dark-shaded histograms are for the more massive
candidates ($m_2\sin{i}$\,$>$\,M$_{\rm up}$).  }
\end{figure*}

A second paper in this series \citep[][ Paper\,II]{Santos-2003} deals
with the metallicity of the stars harbouring planets, providing
accurate spectroscopic parameters for the most recent candidates,
confirming known global properties and also pointing out new emerging
features.  In a third paper, \citet[][ Paper\,III]{Eggenberger-2003}
examine the effect of binarity on planet formation and evolution. The
presence of a close stellar companion is found to bring more massive
planets close-in, in agreement with simulations by \citet{Kley-2001}.

\section{Observational emerging properties}

\subsection{No massive planets on short-period orbits}

In a recent discussion of the statistical properties of massive
planets versus lighter ones \citep{Udry-2002}, we emphasized a
shortage of massive planets on short-period orbits, based on the
sample of about 80 exoplanets known at that time. This feature was
also simultaneously pointed out by \citet{Zucker-2002}. These authors
furthermore verified its statistical significance and very
interestingly examined the possible influence of binarity on the
mass-period relation of exoplanets, an indication of potential
different formation and evolution processes for planets in binaries
and planets around single stars.

With now more than 100 very low-mass candidates (with
$m_2\sin{i}\leq$\,18\,M$_{\rm Jup}$), the lack of short-period massive
planetary companions becomes even clearer, as seen in
Fig.\,\ref{fig1}.
When we neglect the multiple-star systems\footnote{As determined by
  dedicated adaptive optics programmes or unveiled by previous
  spectroscopic measurements \citep[see Paper\,III;][ for a
  review]{Udry-2003:b,Udry-2003:c}} (upper-right panel) in which the
planetary formation or evolution could follow different paths
\citep[][ Paper\,III]{Zucker-2002,Udry-2003:b,Udry-2003:c}, a complete
void of candidates is observed in the diagram for masses larger than
$\sim$\,2\,M$_{\rm Jup}$ and periods smaller than
$\sim$\,100\,days\footnote{The 4 known candidates of planets orbiting
  an evolved star have also been discarded. They however have no
  influence on the results of this study}.  The only remaining point
is {\small HD}\,168443\,b, member of a possible multi brown-dwarf
system \citep{Marcy-2001,Udry-2002}.

Several processes have been proposed to explain the lack of massive
planets on short-period orbits. In the context of the migration
scenario, they mainly follow two different approaches: i) type\,II
migration (after a gap opens in the disk) is shown to be less
effective for massive planets i.e. massive planets stay farther out
than lighter ones, or ii) when the planet reaches the central regions,
some process related to planet-star interactions provokes mass
transfer from the planet to the star - decreasing the mass of the
former - or leads massive planets to fall into the central star. These
approaches will be discussed in more detail in Sect.~\ref{sect3}.

The possibility that multi-planet chaotic interactions send the
lighter candidates in the inner regions (or out) of the system whereas
the massive ones stay in the outer part may also be invoked
\citep{Rasio-96:a,Weidenschilling-96}. The frequency of planets ending
very close to the central star seems however to be small
\citep{Ford-2003}.

\begin{figure}[t]
\centerline{\psfig{width=0.8\hsize,file=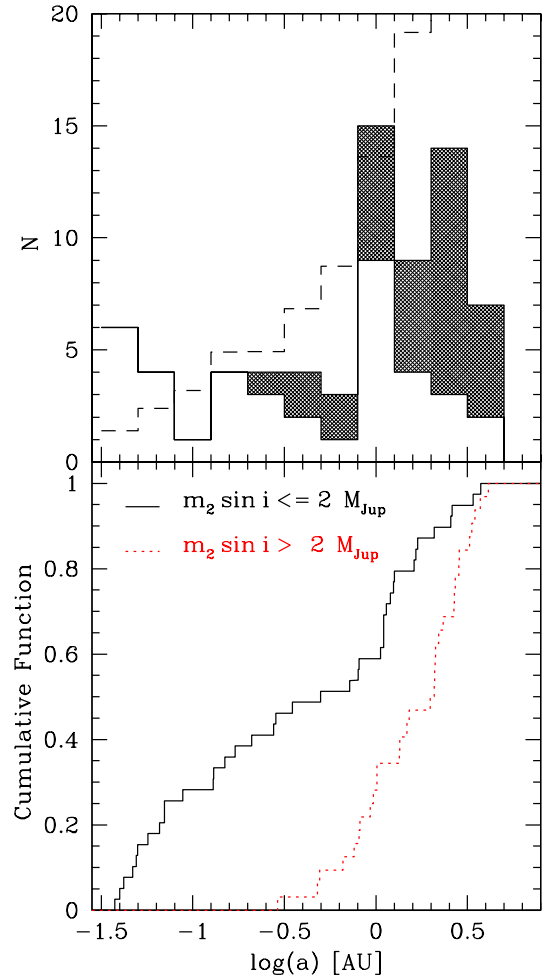}}
\caption{
\label{fig2}
{\sl Upper panel:} Distribution of separations of the known exoplanets
orbiting single dwarf stars (log scale) compared to the similar
distribution of the \citet{Trilling-2002} simulation end-states
(dashed line). The planets more massive than 2\,M$_{\rm Jup}$ are
indicated by the shaded part of the histogram. The planet shortage in
the $0.06\leq a \leq 0.6$\,AU separation range (10--100\,d) comes out
very clearly.  {\sl Lower panel:} Cumulative functions of $\log{a}$
for the 2 mass regimes (limit at 2\,M$_{\rm Jup}$). The preponderance
of light planets close-in and the location of massive planets further
out is clearly emphasized. The Kolmogorov-Smirnov probability for the
2 distributions to come from the same population is $6.7\cdot
10^{-4}$.}
\end{figure}

\subsection{A period valley between 10 and 100-d periods}

In the lower panels of Fig.\,\ref{fig1} and upper panel of
Fig.\,\ref{fig2}, another very interesting feature is coming out of
the period/separation distribution. We observe a shortage of planets
with periods between roughly 10 and 100 days. This observational
property is mainly due to the light candidates
($m_2\sin{i}$\,$\leq$\,2\,M$_{\rm Jup}$).
As seen in the previous section, massive planets orbiting single stars
are almost exclusively found on longer-period orbits. They probably
form and stay further out (see discussion in Sect.\,\ref{sect3}). On
the other hand, lighter planets are found at all distances from their
star\footnote{except the lightest ones ($m_2\sin{i}$\,$\leq$\,$\sim
  0.75$\,M$_{\rm Jup}$) at periods longer than $\sim$\,100~days (see
  next section)}.  This mass-dependent behaviour is also clearly
illustrated by the cumulative functions presented in the lower panel
of Fig.\,\ref{fig2} for two planetary mass regimes.

The features appearing in the planetary period distribution -- peak at
short periods and rise at intermediate periods -- seem significant and
not due to observational biases. The peak at short period is formed by
the pile-up of migrating planets, stopped close to the central star.
As seen in Fig.\,\ref{fig1}, it is almost exclusively composed of the
lowest mass planets ($m_2\sin{i}$\,$\leq$\,0.75\,M$_{\rm Jup}$) and
not due to observational biases as the detection limit at
10\,ms$^{-1}$ is at $P$\,$\simeq$\,4000~days for
$m_2\sin{i}$\,=\,0.75\,M$_{\rm Jup}$ (with $M_1$\,=\,1\,M$_\odot$ and
$e$\,=\,0. The effect of Jupiter on the Sun is 12\,ms$^{-1}$).  The
rise of the distribution at longer periods is now emerging thanks to
the increase of the timebase of the radial-velocity surveys. More and
more long-period planets are being detected. This rise is significant
in the sense that, for a given mass range, similar planets on
shorter-period orbits would have been easier to detect than the actual
ones. The observed {\sl valley} in the period distribution can then
just be seen as a transition region between two categories of planets
which suffered different migration behaviours.

\subsection{A sharp mass transition in the migration process?}

A third important feature emerging from the mass-period diagram is the
apparent lack of very light planets
($m_2\sin{i}$\,$\leq$\,0.75\,M$_{\rm Jup}$) with periods larger than
$\sim$\,100\,days. This feature is already visible in the upper-right
diagram of Fig.\,\ref{fig1} (limited by the dotted line). It becomes
obvious when the planetary minimum mass is displayed with a log scale
(Fig.\,\ref{fig3}).

Of course this feature could be related to the observational bias
inherent to the radial-velocity technique for planet search that makes
the detection more difficult for distant and/or lighter planets. To
obtain a quantitative idea of the effect of this bias, we have also
plotted in Fig.\,\ref{fig3} lines indicating the locations of the
radial-velocity signals (semi-amplitude $K$ of 30, 10 and
3\,ms$^{-1}$), expected on a solar-mass star due to planets on
circular orbits with given minimum masses and semi-major
axes\footnote{The primary mass and eccentricity effects are not
  expected to be important as the radial-velocity semi-amplitude
  scales with $(1-e^2)^{-1/2}M_1^{-2/3}$}.  The line for
$K=10$\,ms$^{-1}$ gives the approximate $3\sigma$-limit of the present
most precise surveys reaching precisions of $\sim$\,3\,ms$^{-1}$
\citep[e.g.][]{Vogt-2000,Queloz-2001:b}.  Some known candidates have
been found very close to this limit, at low mass and small distance or
at high mass and large distance\footnote{We can also note that these
  candidates ``close'' to the detection limit do not present special
  values of the eccentricity and primary or planetary masses}.
However, we clear see an ``intermediate'' region (hatched area),
delimited by $m_2\sin{i}$\,=\,0.75\,M$_{\rm Jup}$, $P$\,=\,100\,d and
$K$\,=\,10\,ms$^{-1}$, where no planetary candidates have been found.
A striking feature of this area is its very sharp limit in mass.

Is this empty region just due to the mentioned observational bias and
the small-number statistics? We have in particular to worry about the
observed effect coming from the shortage of candidates with periods in
the 10--100~day range, as pointed out in the previous section.  So, to
check the statistical significance of the emptiness of the hatched
area in the figure (i.e. to see whether it mainly comes from the small
number of known planets), we performed Monte-Carlo simulations.  The
idea is to obtain $10^6$ realizations of the diagram, reproducing the
actual observational biases of the known sample, and estimate from
them the probability to have no point in the considered zone.

There are basically two ways of producing samples resembling to the
observations: either i) take realistic planetary parameter
distributions and model the observational biases or ii) draw the test
samples from the actually observed distributions in which the acting
biases are built in. As the ``real'' distributions are not known a
priori, we have adopted the second approach:

\begin{figure}[t]
\centerline{\psfig{width=0.9\hsize,file=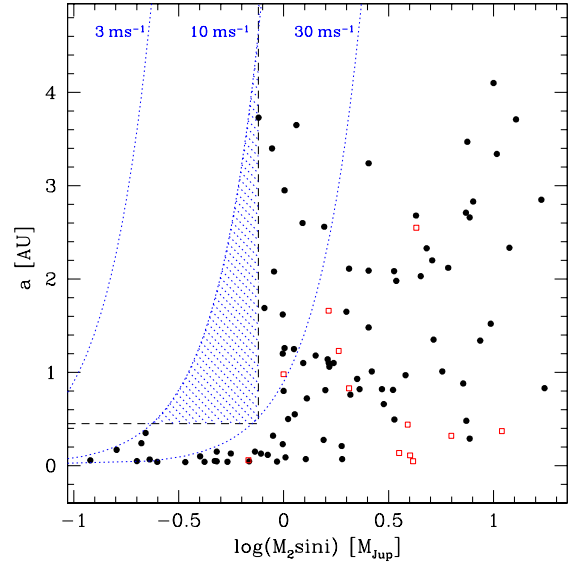}}
\caption{
\label{fig3}
Mass-separation diagram for the known exoplanet candidates. The dotted
lines illustrate the radial-velocity semi-amplitude (3, 10 and
30\,ms$^{-1}$) expected on a solar-mass star due to planets on
circular orbits with given minimum masses and separations. The shaded
area empty of planets is shown not to be due to small number
statistics (see text). Planets in binaries are indicated by open
symbols. }
\end{figure}

i) We randomly select the orbital eccentricities, the primary masses,
the planet minimum masses and separations from the actual observed
distributions of $e$, $M_1$, $m_2\sin{i}$ and $\log{a}$ of the
detected exoplanets. This ensures in particular that we are taking
into account the effect of the orbital eccentricity and of the stellar
mass on planet detection. It also produces distributions that
reproduce the observed {\sl brown-dwarf desert} and the {\sl period
  valley}.
  
ii) Since the distributions of separations are different for {\sl
  light} and {\sl massive} planets (Fig.\,\ref{fig1}), we do the
selection for two mass regimes independently (limit at 4\,M$_{\rm
  Jup}$).

\begin{figure}[t]
\centerline{\psfig{width=0.9\hsize,file=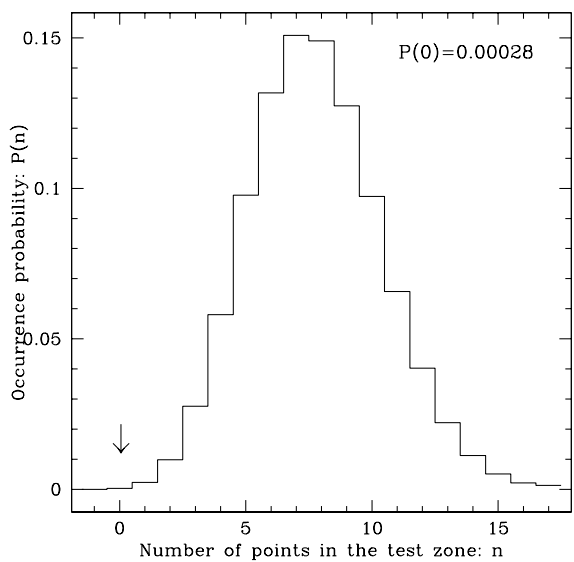}}
\caption{
\label{fig4}
Result on the statistical significance of the hatched zone in
Fig.\,\ref{fig3} based on $10^6$ realizations of the diagram with
Monte-Carlo simulations. The figure presents the normalized histogram
of the distribution of the number of points, $n$, found in the test
region i.e. the occurrence probability of $n$.  The arrow location at
$n=0$ indicates the position of the actual observational result.  Its
probability of occurrence only by chance is $P(0)$\,=\,0.00028.}
\end{figure}

iii) We only use the distributions for planets orbiting single dwarf
stars as 1) planets in binaries may have different characteristics
\citep[][ Paper\,III]{Zucker-2002} and 2) the post main-sequence
stellar evolution may drastically change the fate of short-period
planets.

iv) We reject the events giving a radial-velocity semi-amplitude
smaller than $K_{\rm lim}$\,=\,10\,ms$^{-1}$. They would hardly be
detected by the present surveys. A less restrictive limit a
15\,ms$^{-1}$ is also considered to check the sensitivity of the
results on the detection limit.

v) Each realization consists of a selection of 86 pseudo-planets,
corresponding to the actual observations (64 ``light'' and 22
``massive'' planets orbiting single dwarf stars, as on February 2003).
For each of them we count the number $n$ of points in the ``test
zone''. Note that the detection limit at $K_{\rm lim}$ is no more as
well defined as in Fig.~\ref{fig3} but it now takes into account the
orbital eccentricity and primary-mass values.

\begin{figure}[t]
\centerline{\psfig{width=0.9\hsize,file=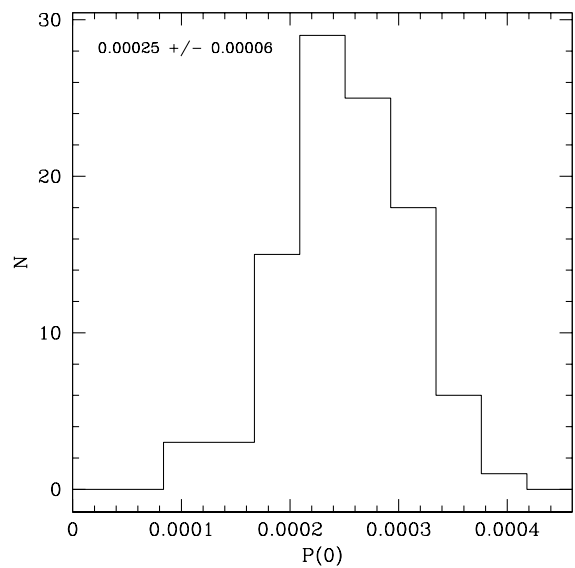}}
\caption{
\label{fig5}
Distribution of the $P(0)$ probabilities of 100 bootstrap simulations
(but with $10^5$ realizations each) conducted to estimate the
uncertainty on the value obtained in Fig.\,\ref{fig4}. They yields a
value 0.00025\,$\pm$\,0.00006 for $P(0)$.}
\end{figure}

The normalized histogram of the distribution of the number of points
$n$ in the test zone from our $10^6$ realizations is shown in
Fig.\,\ref{fig4}. The probability of having no detection is only
0.00028.  A bootstrap of the procedure with 100 times $10^5$
realizations gives a value of 0.00025\,$\pm$\,0.00006
(Fig.\,\ref{fig5}), in agreement with the previous result.  Relaxing
the detection limit to 15\,ms$^{-1}$ yields a probability of 0.004.
It shows the low sensitivity of the result on the chosen
radial-velocity detection limit.  We thus can conclude that the
considered area is indeed empty of planets with a confidence level of
about 99.97\,\%.

This conclusion is only valid if the observational biases are
correctly taken into account. Biases related to planetary parameter
distributions and detection technique are by construction included in
the simulations.  However, other aspects play a role. For example
activity-induced radial-velocity jitter may screen planet detection.
Such an effect can be neglected if we suppose that the $m_2\sin{i}$,
$e$, $M_1$ and $a$ parameters are not correlated with activity.
Another concern relates to the difficulty encountered when trying to
actually derive the values for the orbital elements, what is different
from just detecting radial-velocity variability.  Observation timing
and phase coverage are then also important parameters. These remarks
emphasize the caution needed when considering the above described
results. We should see them more like a trend emerging from the data
rather than a real proof.

In the next section, we will propose a tentative explanation for this
feature, based on recent results on runaway migration studied by
\citet{Masset-2003}. We will also discuss in the last section of this
paper the implications on the radial-velocity planet-search survey of
the paucity of light planets with intermediate periods.

\section{Checking the theoretical approaches}
\label{sect3}

The different features pointed out above in the various presentations
of the period/separation--mass diagram will allow us to check the
processes proposed to explain the observed distribution of periods for
extra-solar planet candidates.

\subsection{Low migration rate for massive planets}

A first approach to explain the lack of massive planets on
short-period orbits invokes a low migration efficiency for higher mass
planets. This view is supported by recent simulations of single-planet
migration. When the mass of the planet becomes of the order of a
significant fraction of the characteristic disc mass with which it
interacts, the inertia of the planet becomes important and slows down
the orbital evolution \citep[e.g.][]{Trilling-98,Nelson-2000}. It also
takes more time to open a gap in the disk for massive planets -
eventually longer than the lifetime of the disk - and initiate
type\,II migration \citep{Trilling-2002}.

Using a model assuming a simple impulse approximation for the type\,II
migration, no {\sl ad-hoc} stopping mechanism in the center and
neglecting type\,I migration\footnote{The extremely short timescale of
  the type\,I migration represents a serious theoretical problem for
  planet formation. However, suggestions have been recently made that
  co-orbital corotation torque in thin viscous disks may decrease its
  dramatic effect \citep{Artymowicz-2003} and even stop or revert it
  for a large set of conditions \citep{Masset-2002}. This has however
  to be further examined},
\citet{Trilling-2002} show that, for identical initial disk
conditions, close-in surviving pseudo-planets have smaller masses and
distant surviving pseudo-planets have larger masses. They also find
that planets forming farther out migrate less rapidly and that
higher-mass planets migrate on longer timescales.  Thus, there should
be more massive planets at intermediate and large semi-major axes than
found close-in.  On the contrary, the population of short-period
planets should be dominated by smaller-mass planets, what is indeed
observed:

-- The upper panel in Fig.\,\ref{fig2} presents the logarithmic
distribution of exoplanet separations, superimposed to the
\citet{Trilling-2002} results. As already pointed out by these authors,
if we take into account the
observational bias penalizing long-period planets in the
radial-velocity surveys, the theoretical and observational results
agree fairly well.  The observed peak at small separations is not
reproduced by the simulations of  Trilling et al. because they have
not introduced in their study any {\sl ad-hoc} process to stop the
migration close to the central star.

-- Furthermore, the lower panel in Fig.\,\ref{fig2} giving the
cumulative functions for 2 mass regimes (limit at 2\,M$_{\rm Jup}$)
clearly shows that low-mass planets form the predominant population of
short-period orbits whereas more massive planets are located further
out.

\begin{figure}[t!]
\centerline{\psfig{width=0.9\hsize,file=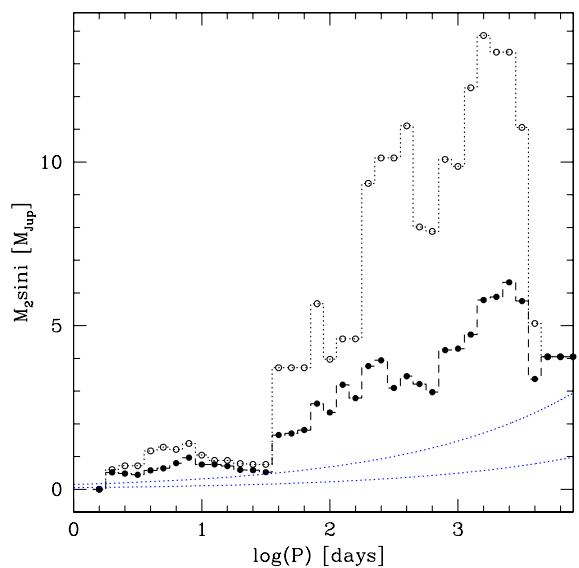}}
\caption{
\label{fig6}
Mean mass (filled circle) or highest mass (average on the 3 highest
values; open circles) of planets 
in period smoothing windows of width $\log{P[{\rm days}]}$\,=\,0.2. A
clear increase of planet maximum masses with period is observed,
despite the fact that massive planets observed at a given period
should be easier to detect at smaller periods.  Detection limits at 10
and 30\,ms$^{-1}$ ($M_1$\,=\,1\,M$_\odot$, $e$\,=\,0) are represented
by the dotted lines.}
\end{figure}

-- Moreover, the lower panels of Fig.\,\ref{fig1} indicates that you
have to go further and further out to find more and more massive
planets. The hatched histogram representing planets more massive than
4\,M$_{\rm Jup}$ rises up to 1000\,days (left panel) -- although such
massive planets should be easier to detect at smaller periods --
whereas the maximum of the rise of the distribution of lighter-mass
planets is around 400\,days.  If we decrease down to 2\,M$_{\rm Jup}$
the limit between the 2 mass regimes, the mentioned features almost
overlap (right panel), indicating a trend for the higher masses to be
found at larger distances from the central star.  This higher mass
trend with period is explicited in Fig.\,\ref{fig6} which displays the
mean mass (filled circle) or higher mass (average on the 3 highest
values; open circles) of planets
in period smoothing windows of width $\log{P[{\rm days}]}$\,=\,0.2.
The effect becomes visible around $P$\,=\,40\,d and turns to be very
effective around $P$\,=\,100\,d. Detection limits at 10 and
30\,ms$^{-1}$ are represented by the dotted lines. They show that
observational biases related to the radial-velocity technique are not
responsible for the observed feature.  This effect is not surprising
considering first that massive planets are preferentially formed in
the outer regions where a larger amount of building material is
available and second that higher-mass objects migrate on longer
timescales.  Furthermore, as the disk dissipates, the migration of
more massive planets slows down earlier since their larger angular
momentum produces greater resistance to disk-induced migration
\citep{Armitage-2002}.

\subsection{Runaway migration and lack of light planets with
  longer periods}

A recent study by \citet{Masset-2003} brings some insights for our
understanding of the observed lack of small-mass planets
($m_2\sin{i}$\,$\leq$\,0.75\,M$_{\rm Jup}$) on intermediate-period
orbits ($P$\,$\geq$\,100\,d). These authors studied the effect of
co-orbital corotation torque on migrating protoplanets in the case of
massive protoplanetary disks for which the planet Hill radius and the
disk thickness have comparable orders of magnitude.
In particular they show that, if the mass deficit created by the
radial drift with the planet of the material trapped in the co-orbital
region is larger than the planet mass, the migration rate undergoes a
runaway which can rapidly vary the protoplanet semi-major axis by a
large amount (50\% over a few tens of orbits).  The authors state
that: {\sl ``this can happen only if the planet mass is sufficient to
  create a gap in its surrounding region and if the surrounding disk
  mass is larger than the planet mass. This typically corresponds to
  planet masses in the sub-Saturnian to Jovian mass range embedded in
  massive protoplanetary disks''.}

Moreover, the simulations by \citet{Masset-2003} show that the limit
in planetary mass for the appearance of the runaway regime is very
steep (their Figs.\,12-14).  The threshold conditions for runaway may
vary widely depending on the disk mass, viscosity or thickness. It
however depends relatively weakly upon the planet mass on the
high-mass side i.e. their is a critical mass relatively
well-constrained ($M_{\rm crit}$\,$\simeq$\,1\,M$_{\rm Jup}$) under
which runaway is likely provided that the protoplanetary disk is not
too light weight, and above which runaway is impossible (Masset, priv.
comm.). As our observed limit at $m_2\sin{i}$\,$\simeq$\,0.75\,M$_{\rm
  Jup}$ is compatible with this critical mass, there could well be a
link between the statistical observed property and runaway migration.

In this picture, planets which happen to be massive enough
($M$\,$>$\,$M_{\rm crit}$) before reaching the central regions would
end up as the scattered population on the right of Fig.\,\ref{fig3},
whereas planets that never grow enough would remain very mobile and
would eventually be flushed into the very central region.

Note that another limit exist for runaway migration on the planet
low-mass side. Thus, very light planets (typically much lighter than
Saturn) do not encounter such type of fast migration. That can be the
case for the outer icy planets of the Solar System.

\subsection{Observed constraints for migration: summary}

In summary, putting everything together, we can try to draw a global
picture for the planet migration behaviour in function of the
planet mass, in single-star systems:\\
\indent i) The more massive planets (typically, masses larger than
$\sim$\,4\,M$_{\rm Jup}$) form preferentially in the outer regions --
where there is a large-enough material reservoir -- and do not migrate
much. None is observed within $\sim$\,0.5\,AU from the central star.\\
\indent ii) Intermediate-mass objects migrate more easily whatever the
distance where they form, the migration rate depending on the local
conditions (planet mass, disk mass, viscosity, etc). They are observed
at all distances.\\
\indent iii) The lighter planets (masses from sub-Saturnian to Jovian)
migrate easily, possibly undergoing runaway migration bringing them in
the central tenth of an AU from the star if the disk is massive enough.
Most of them are actually found in the very central regions.  None is
observed with $P$\,$\geq$\,$\sim$\,100\,days although radial-velocity
surveys are now precise enough to detect them.\\
\indent iv) We know, however, from our own Solar System that this is
not true for the planets that are far away or much lighter than the
known exoplanets.  Thus, we can speculate that for the very light
planets the migration efficiency is decreasing with distance or mass.
This again fits into the the runaway migration scenario, the
simulations showing the existence of a transition limit in mass, on
the low-mass side, for the runaway to occur. Unfortunately, the
precision achieved by the present surveys does not allow us to
determine the separation or mass limits at which the migration slows
down significantly. Good hope to answer this question is however
brought by the future surveys at higher precision \citep[like {\small
  HARPS} at 1\,ms$^{-1}$,][]{Pepe-2002:b} that will increase our
detection sensibility by a large factor (curve at $K$\,=\,3\,ms$^{-1}$
in Fig.\,\ref{fig3}).

This {\sl observational} view of planet migration may be changed by
the influence of a perturbing stellar companion (Paper\,III) that will
make the massive planets migrate inwards faster or make closer-in
intermediate-mass planets grow bigger \citep{Kley-2001}.

Finally, we have to stress that although theory suggests that it is
possible to form massive planets closer to the stars than the
ice-boundary limit
\citep{Papaloizou-99,Bodenheimer-2000,Sasselov-2000}, within the
migration scenario, there is no observational requirement for planets
to form at separations smaller than a few AU.

\subsection{Central star-planet interaction}
\label{sect3.4}

From the {\sl central star -- planet} interaction point of view,
several scenarios were proposed to explain the lack of massive planets
on short-period orbits as alternatives to the above developed
arguments in the context of the migration scenario.

\citet{Patzold-2002} proposed to explain the lack of massive close-in
planets by tidal interactions between the planet and its central star.
If the planet orbital period is smaller than the stellar rotation
period, tidal friction will spin up the star whereas the semi-major
axis of the planetary orbit is decreased, eventually reaching the
Roche zone of the central star on a short timescale.  These authors
pointed out that massive planets spiral in much faster than others,
explaining thus the observed trend.

\citet{Trilling-98} had already addressed this problem in also
considering mass transfer between the planet and the star. Some of the
pseudo-planets in their simulations reach the Roche lobe limit, start
to lose mass and thus experience an outward torque from the Roche
lobe overflow.  They estimate the maximum mass of surviving planets to
be around 2\,M$_{\rm Jup}$, corresponding to what is indeed observed.

Another approach could relate to the evaporation of the planet
atmosphere when the planet comes close to the central star, decreasing
thus its mass. Up to now, studies based on the $T_{\rm eff}$ of the
heated close-in planets have shown that hydrogen evaporation was not
efficient in the case of the known Hot Jupiters
\citep[e.g.][]{Mayor-2000:a}.  However, new developments based on the
estimate of the exospheric temperature of {\object HD\,209458\,b},
evaluated to be much higher than $T_{\rm eff}$
\citep{Vidalmadjar-2003}, call for a complete reconsideration of the
question.


These approaches are very interesting to explain the period
distribution of planets very close to the central stars
($a$\,$\leq$\,0.1\,AU, with a pile up around 3 days), especially
considering that the peak at very short periods is almost entirely
formed by the lightest mass planets
($m_2\sin{i}$\,$\leq$\,0.75\,M$_{\rm Jup}$).  However, they do not
explain the extended range of periods ($P$\,$\leq$\,100\,d) in which
no planets more massive than $\sim$\,2\,M$_{\rm Jup}$ are found.  This
large interval supports the non-migration of massive companions rather
than the disappearance of close-in migrating massive planets.

\section{Summary and concluding remarks}

To summarise the main observational features pointed out in
this paper and their main implications, we have that:

1) No massive planets ($m_2\sin{i}$\,$\ge$\,$\sim$\,2\,M$_{\rm Jup}$)
are found on short-period orbits ($P$\,$<$\,$\sim$\,100\,d) around
single stars. This is not an observational bias as these candidates
are the easiest ones to detect.

2) The maximum mass of detected planets per period interval 
increases with distance to the central star (Fig.\,\ref{fig6}). This
is also a solid result as massive planets found at a given distance
are easier to detect closer in.

3) The above points 1) and 2) suggest that the migration rate of
planets decreases with increasing mass of the planetary
companion. This result agrees with recent simulations of migrating
planets in viscous disks
\citep{Trilling-98,Trilling-2002,Nelson-2000}.  We thus expect a large
number of massive planets to be on long-period orbits and so to be
still undetected because of the limited duration of the present surveys.
A large number a lower-mass planets probably also exist on long-period
orbits; they are however more difficult to detect. These
represent primary targets for future higher-precision surveys
\citep[like e.g. {\small HARPS},][]{Pepe-2002:b} whereas the youngest
among the formers are interesting targets for direct imaging of
planetary-type objects. This result is also very important for
on-going radial-velocity planet searches from which we now expect an
increasing number of planet candidates as the survey durations will
grow. The ``older'' the survey, the higher the expected detection
rate. In this context, the first epoch measurement of the targets is
an important parameter for planet detection.

4) We observe a shortage of planets with periods in the 10--100\,d
range. This {\sl valley} in the distribution is located just
in-between the peak of light planets that have migrated inwards and
were stopped close to the central star and the rise of the
distribution due to the increasing number of detected planets with
longer periods.

5) Up to now, no planet candidates with very low masses
($m_2\sin{i}$\,$\leq$\,0.75\,M$_{\rm Jup}$) have been detected on
orbits with periods longer than $\sim$\,100\,days.  This trend seems
significant and not due to small-number statistics or to the
particular shape of the period or separation distributions. The limit
is sharp in mass, indicating a strong transition in the migration
process for different mass regimes. These features are in agreement
with predictions of runaway migration simulations of planets lighter
than $\sim$\,1\,M$_{\rm Jup}$ in massive protoplanetary disks
\citep{Masset-2003}.

Note however that this lack of light planets on intermediate periods
may not be confirmed for much lighter planets or for planets in the
considered mass range but formed far out from the central star (e.g.
Neptune- and Uranus-type planets), that are still out of reach of the
present observational facilities and detection techniques.
 
In conclusion, we have pointed out a clear dependence of migration
on mass and distance from the central star, providing important
constraints for models of planetary formation (migration). More
quantitative constraints for the different mass regimes will require
a better detection capability of the surveys, to diminish the
effect of the observational bias in the distributions and thus
increase the confidence we can put in the derived trends, especially
for the lower-mass planets.

\begin{acknowledgements}
  This study has benefited from fruitful discussions with Frederic
  Masset and Willy Benz.  We thank the Geneva University and the Swiss
  NSF (FNRS) for their continuous support of our planet-search
  projects.  Support to N.S.  from Funda\c{c}\~ao para a Ci\^encia e
  Tecnologia (Portugal) in the form of scholarships is gratefully
  acknowledged.
\end{acknowledgements}


\bibliographystyle{aa} 
\bibliography{udry_articles}

\end{document}